\documentclass[prl,twocolumn,superscriptaddress,showpacs]{revtex4}

\usepackage{graphicx}

\def\be{\begin{equation}}
\def\ee{\end{equation}}

\begin{document}

\title{Crack Path Prediction in Anisotropic Brittle Materials} 

\author{Vincent Hakim$^1$ and Alain Karma\footnote{
Permanent address: Physics Department and Center 
for Interdisciplinary Research on 
Complex Systems, Northeastern University, 
Boston, Massachusetts 02115}}
 
\affiliation{Laboratoire de Physique Statistique,
Ecole Normale Sup\'erieure, 24 rue Lhomond, 75231 Paris, France}

\date{\today}

\begin{abstract}

A force balance condition to predict 
quasistatic
crack paths in anisotropic brittle materials
is derived from an analysis of diffuse interface continuum
models that describe both short-scale failure inside a
microscopic process zone 
and macroscopic linear elasticity. The derivation exploits
the gradient dynamics and translation symmetry properties of this class 
of models to define a generalized energy-momentum tensor whose integral
around an arbitrary closed path enclosing
the crack tip yields all forces acting on this tip,
including Eshelby's configurational forces, cohesive forces,
and dissipative forces. This condition is 
validated quantitatively by numerical simulations.
\end{abstract}
\pacs{62.20.Mk, 46.50.+a, 46.15.-x}

\maketitle

The prediction of the path chosen by a crack  
as it propagates into a brittle material
has been a long standing problem in fracture 
mechanics. This question has been addressed primarily
in a theoretical framework where the equations
of linear elasticity are solved with zero traction 
boundary conditions on crack surfaces 
that extend to a sharp tip \cite{refgen}. The stress distributions 
near the tip have universal divergent forms
\be
\sigma_{ij}^m(r,\Theta)=\frac{K_m}{\sqrt{2\pi r}}f_{ij}^m(\Theta), 
\label{km}
\ee
where $K_m$ are the stress intensity factors
for the three standard modes I, II, or III of fracture
($m=1,2$ or $3$) and $\Theta$ 
is the angle between the radial vector of
magnitude $r$ with
origin at the crack tip and the local   
crack direction.
For the crack to propagate, 
the energy release rate (or crack
extension force)
\be
G=\alpha(K_1^2+K_2^2)+K_3^2/(2\mu),\label{Gdef}
\ee
must exceed some material dependent threshold $G_c$
that is theoretically equal to twice the surface energy ($G_c=2\gamma$),
but often larger in practice. Here, $\nu$ is Poisson's ratio,
$E$ is the bulk modulus, $\mu$ is the shear modulus,
and $\alpha\equiv (1-\nu^2)/E$.

Like other problems in fracture, 
the prediction of crack paths
was first examined \cite{BarChe1961} for mode III  
which is simpler because
the antiplane
component of the displacement vector
$u_3$ is a purely scalar Laplacian field.
The stress distribution
near the tip, can be expanded as  
\begin{equation}
\sigma_{3\Theta}\equiv  \frac{\mu}{r}
\frac{\partial u_3}{\partial \Theta}
=\frac{K_3}{\sqrt{2\pi r}}\cos 
\frac{\Theta}{2}-\mu A_2 \sin \Theta
+\dots, \label{sigma3}
\end{equation}
and the dominant divergent contribution is always symmetrical 
about the crack direction. This implies that
knowledge of $K_3$
cannot predict any other path than a straight one. 
To avoid this impasse, 
Barenblatt and Cherepanov \cite{BarChe1961} retained the
subdominant term $\sim \sin\Theta$, which breaks this symmetry, and
hypothesized that a curvilinear crack
propagates along a direction 
where $A_2=0$, and hence when the stress distribution
is \emph{symmetrical} about the crack direction. 
In subsequent extensions of this work, 
several criteria have been proposed
for plane loading, for which 
the tensorial nature
of the stress fields 
makes it possible to predict non-trivial crack
paths purely from the knowledge of the stress-intensity
factors
\cite{GolSal1974, CotRic1980}.
The generally-accepted condition ``$K_2=0$'' 
assumes that the crack propagates
in a pure opening mode with a symmetrical stress
distribution about its local axis \cite{GolSal1974} and
is the direct analog  
for plane strain ($u_3=0$) of the condition 
$A_2=0$ for mode III.
This ``principle of local symmetry'' has been  
rationalized using plausible arguments \cite{CotRic1980} but 
cannot be derived  
without an explicit description of the
process zone, where elastic strain energy is both dissipated
and transformed nonlinearly into
new fracture surfaces. As a result, how to extend
this principle to anisotropic materials, where  
symmetry considerations have no obvious generalization,
is not clear \cite{Mar2004}. In addition, path
prediction remains largely 
unexplored for mode III even 
for isotropic materials. 

In this letter, we address the problem of path
prediction in the context of continuum models of
brittle fracture that describe both short scale failure
and macroscopic linear elasticity within a 
self-consistent set of equations. 
Such models have already proven capable to reproduce a wide range of
phenomena for both antiplane \cite{KarLob2004} and plane 
\cite{HenLev2004} loading from the onset of crack propagation at
the Griffith threshold to dynamical 
branching instabilities \cite{KarLob2004}
and oscillatory \cite{HenLev2004} instabilities.
From an analysis of these models, we derive a new condition
to predict crack paths that is interpreted physically in
the context of previous results from the fracture community.

For clarity of exposition, 
we base our derivation on the
phase-field approach of Ref. \cite{KKL2001} where the displacement
field is coupled to a single
scalar order parameter or ``phase field'' $\phi$,
which describes a smooth transition in space between unbroken 
($\phi=1$) and broken
states ($\phi=0$) of the material. Our approach is sufficiently
general, however, to be applicable to 
a large class of diffuse interface descriptions 
of brittle fracture. We focus on 
quasi-static fracture in a macroscopically isotropic elastic
medium with negligible inertial
effects. Material anisotropy is simply included
by making the surface energy, $\gamma(\theta)$, dependent on
the orientation $\theta$ of the crack direction with
respect to some underlying crystal axis.  

For   
brevity of notation, we define the four-dimensional vector field  
$\psi^k=u_k$ for $1\le k \le 3$ and
$\psi^4=\phi$  
where $u_k$ are the components
of the standard displacement field.
The energy density ${\cal E}$ 
depends on $\phi$ and  
$\partial_j\psi^k \equiv \partial \psi^k/\partial x_j$, where
spatial gradients of the displacement contribute 
to the elastic strain energy and
gradients of the phase-field contribute to the surface energy
\cite{KKL2001}. The equations of motion 
are derived variationally from the spatial integral of $\cal{E}$, the
total energy $E$ 
of the system,
and obey the gradient dynamics
\be 
 \delta_{k,4}\chi^{-1}\partial_t\phi=
-\frac{\delta E}{\delta \psi^k}=\partial_j
\frac{\partial {\cal E}}{\partial
\partial_j \psi^k}-\frac{\partial {\cal E}}{\partial \psi^k},\label{eqmo}
\ee
where $\delta_{i,i}=1$ and $\delta_{i,j}=0$ for $i\ne j$.
These Euler-Lagrange equations for $\psi^k=u_k$ are
simply the static equilibrium conditions 
that the sum 
of all forces 
on any material
element 
vanish. 
The fourth equation for $\psi^4=\phi$
is the standard Ginzburg-Landau form that governs
the phase-field evolution, where
$\chi$ is a kinetic coefficient that controls the rate
of energy dissipation in the process zone. 

In the present model that describes both microscopic
and macroscopic scales, the problem of 
predicting the macroscopic path of a crack 
can be posed as an ``inner-outer'' matching problem.
We seek inner solutions of the equation of motion (\ref{eqmo})
on the scale $\xi$ of the process zone subject to
far field boundary conditions imposed by matching these
solutions to the standard solutions of linear elasticity on the
outer scale of the system $W\gg \xi$. These outer solutions change
slowly on a scale where the crack advances by a distance 
$\sim \xi$. We can therefore search for inner solutions
by rewriting Eq.~(\ref{eqmo})  
in a frame translating uniformly 
at the instantaneous crack speed $V$
parallel to the crack direction 
\be 
- \delta_{k,4} V\chi^{-1} \partial_1 \phi =
\partial_j
\frac{\partial {\cal E}}{\partial
\partial_j \psi^k}-\frac{\partial {\cal E}}{\partial \psi^k}.\label{eqmo2}
\ee
We then seek solutions of Eq.~(\ref{eqmo2}) with far 
displacement fields ($r\gg \xi$) in the unbroken solid 
that yield the singular stress distributions
defined by Eqs.~(\ref{km}) and (\ref{sigma3}). 

There are  
two ways to proceed to solve
this problem. The first approach, which will be exposed
in more details elsewhere, exploits the existence of
the stationary semi-infinite
crack solution, $\psi^k_0$, for $K_2=A_2=0$ and $G=G_c$
which is symmetrical about the crack axis. 
One can then seek solutions of Eq.~(\ref{eqmo2})  
linearized around this Griffith crack 
with the driving force for crack advance 
$G-G_c$ and symmetry breaking
perturbations ($K_2$, $A_2$, and the anisotropy 
of the surface energy $\partial_\theta \gamma$) 
assumed to be small. Owing to the
variational character of the phase-field
equations, the linear operator of the resulting linearized problem
is self-adjoint and hence has two zero modes,
$\partial_i\psi^k_0$, associated with translations of the 
Griffith crack. 
The standard requirement that non-trivial solutions 
to this problem be orthogonal to the null space of
the adjoint linear operator yields two independent solvability
conditions (for $i=1,2$).

The second approach, which we adopt here, 
exploits the variational
character of the equations of motion. 
It yields identical solvability conditions
as the first approach when $G-G_c$ and
symmetry breaking perturbations are small, but it is
more general since it does not require
these quantities to be small.  
We start from the equality  
obtained simply from chain rule differentiation,  
\be
\int_{\Omega} d\vec x~  \partial_i {\cal E}
=\int_{\Omega} d\vec x ~\left(\frac{\partial {\cal E}}{\partial 
\psi^k}\partial_i\psi^k 
+\frac{\partial{\cal E}}{\partial \partial_j\psi^k}
\partial_j\partial_i\psi^k \right),\label{chain}
\ee
where $d\vec x\equiv dx_1dx_2$ and $\Omega$ is an arbitrary region of the
$(x_1,x_2)$ plane.
Using Eq.~(\ref{eqmo2}) to eliminate $\partial {\cal E}/\partial \psi_k$
from the integrand of the right-hand-side (r.h.s.), 
we obtain 
\be
F_i\equiv \int_{\Omega} d\vec x ~\partial_j\,T_{ij}-\frac{V}{\chi}
\int d\vec x~\partial_1\phi\partial_i \phi =0
~~{\rm for}~~i=1,2.\label{div}
\ee
The generalized 
energy-momentum (GEM) tensor
\be
T_{ij}\equiv {\cal E}\delta_{ij}
-\frac{\partial{\cal E}}{\partial \partial_j\psi^k}
\partial_i \psi^k 
\ee
extends the classical energy-momentum tensor
of linear elastic fields \cite{Esh1975} by 
incorporating
short-scale physics through its additional dependence
on the phase-field.

\begin{figure}[ht]
\includegraphics[width=5cm]{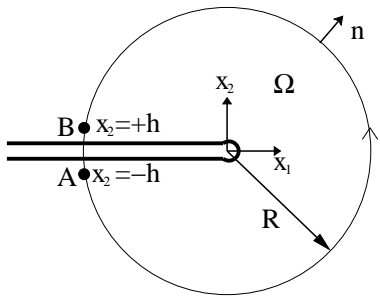}
\caption{Spatially diffuse crack tip region with $\phi=1/2$
contour separating broken and unbroken material (thick solid line).
}
\label{schem}
\end{figure}

We now consider a region $\Omega$ 
that contains
the process zone (crack tip) and
write
the integral of the divergence
of the GEM tensor 
as a contour integral 
\be
F_i=\int_{A\rightarrow B}\!\!\!\!\!\!\!\!\!\! ds \,T_{ij}\,n_j
+\int_{B\rightarrow A}\!\!\!\!\!\!\!\!\!\! ds \,T_{ij}\,n_j -\frac{V}{\chi}
\int_\Omega d\vec x~\partial_1\phi\partial_i \phi=0.\label{line}
\ee
We have decomposed the boundary of $\Omega$
into: (i) a
large loop $(A\rightarrow B)$ around the tip in the unbroken
material, where $A$ ($B$) is at a height $h$ below (above)
the crack axis that is much larger than the process zone size but
much smaller than the radius $R$ of the contour, $\xi \ll h \ll R$,
and (ii)
the segment $(B\rightarrow A)$
that traverses the crack from $B$ to $A$ behind the tip, 
as illustrated in Fig.~\ref{schem}. In both integrals,
$ds$ is the contour arclength element  
and
$n_j$ the components of its outward normal.

Eq.~(\ref{line}) provides the basis to predict the crack
speed and its path for quasi-static fracture. The 
$F_i$ can be interpreted as the
sum of all forces acting on the crack tip 
parallel ($i=1$) and perpendicular ($i=2$)
to the crack direction, including configurational,
cohesive, and dissipative forces,   
represented by the three integrals terms from left to right,
respectively.
In the unbroken material where $\phi$ is constant,
the tensor $T_{ij}$ reduces identically to the
energy-momentum tensor introduced by Eshelby to compute the 
configurational force on the crack tip 
treated as a defect in a linear elastic
field \cite{Esh1975}, and used in subsequent
attempts to derive criteria
for crack propagation and stability 
\cite{Guretal,Adaetal1999,Ole2003}. 
Thus, the first integral in 
Eq.~(\ref{line}) yields the  
configurational forces 
\be
\int_{A\rightarrow B} ds \,T_{1j}\,n_j=G,\ 
\int_{A\rightarrow B} ds \,T_{2j}\,n_j=G_\theta(0),  \label{f2}
\ee
in the limit $h/R\rightarrow 0$.
The first is the crack extension
force and also Rice's $J$ integral \cite{Ric1968}.
The second is the Eshelby torque
$G_\theta\equiv dG(\theta)/d\theta$ \cite{Esh1975}, where
$G(\theta)$ is the extension force at the tip
of a crack extended at a vanishingly small angle $\theta$ from
its local direction.  This 
torque tends to turn the crack in a direction
that maximizes $G$.

An important new ingredient of the present derivation is
the second portion of the line integral ($\int_{B\rightarrow A}$) 
of the GEM tensor that traverses the crack. 
This integral represents physically the
contributions of cohesive forces inside the
process zone. To see this, we first note that 
the profiles of the phase-field and the three components of the
displacement can be made to
depend only on $x_2$ provided that the contour is chosen much larger
than the process zone size and to traverse the crack perpendicularly
from $B$ to $A$. With this choice, we have that $n_1=-1$,  
$n_2=0$, 
along this contour and therefore 
that, for $i=1$
\be
\int_{B\rightarrow A} ds \,T_{1j}\,n_j=
-\int_{-h}^{+h} dx_2 T_{11} = -2\gamma,\label{surf}
\ee
where the second equality follows from the fact that
spatial gradients parallel to the crack direction ($\partial_1\psi^k$) 
give vanishingly small contributions
in the limit $h/\xi\rightarrow +\infty$ and
$R/\xi\rightarrow +\infty$ with $h/R\rightarrow 0$. 
This yields the expected result that
cohesive forces exert a force opposite to the crack extension force
with a magnitude equal to twice the surface energy. An analogous
calculation for $i=2$ yields, in the same limit $\xi \ll h \ll R$,
the other component of the force
perpendicular to the crack direction  
\be
\int_{B\rightarrow A} ds \,T_{2j}\,n_j=
-\int_{-h}^{+h} dx_2 T_{21}= 
-2  \gamma_\theta(0).\label{herring}
\ee
This force is the direct analog of the Herring torque 
$\gamma_\theta=d\gamma/d\theta$ on
grain boundaries \cite{Her1951}. This torque tends
to turn the crack into a direction that minimizes the
surface energy. Substituting the results 
of Eqs.~(\ref{f2}) to (\ref{herring})
into Eq.~(\ref{line}), we obtain the two conditions
\begin{eqnarray}
F_1&=&G-G_c-f_1=0,\label{x1}\\
F_2&=& 
G_\theta(0) 
 -G_{c\theta}(0) 
-f_2=0,\label{x2}
\end{eqnarray} 
where we have used the fact
that $G_{c\theta}=2\gamma_{\theta}$, and 
defined the dissipative forces 
\be
f_i=V \chi^{-1}\int_{-\infty}^{+\infty}\int_{-\infty}^{+\infty}
dx_1dx_2
~\partial_1\phi\partial_i\phi, \label{dissforces}
\ee
by letting the area $\Omega$ tend to infinity
since the integrand vanishes 
outside the process zone. Eq.~(\ref{x1}) predicts the
crack speed $V\approx \chi(G-G_c)/\int d\vec x (\partial_1\phi_0)^2$ 
for $G$ close to $G_c$ where $\phi_0$ is the phase-field
profile for a stationary crack \cite{KarLob2004}.
Eq.~(\ref{x2}), in turn, predicts the crack path.
$G$ and $G_\theta$ can be 
generally 
obtained from
Eq.~(\ref{f2}),
using the known forms of the
displacement fields near the tip.
The $J$ integral yields Eq.~(\ref{Gdef}) for $G$.  
$G_\theta$ can also be obtained directly from the
expression for $G(\theta)$. The latter   
is instructive here to highlight important differences
between plane strain and antiplane shear. Consider a
straight crack parallel to the $x_1$ axis
with stress intensity factors $K_1$
and $K_2$. Now extend this crack by a length $L$
at a small angle $\theta$ from this axis.
The new stress intensity factors  
are given by
$K_1^*\approx  K_1-3K_2\theta/2$ and
$K_2^*\approx K_2+K_1\theta/2$ to linear order
in $\theta$ \cite{AmeLeb1992}
independent of $L$. Using Eq.~(\ref{Gdef}) with
these new stress intensity factors to define
$G(\theta)$, we obtain at once $G_\theta(0)=-2\alpha K_1K_2$.
Substitution 
in Eq.~(\ref{x2}),
provides the condition  
\be
K_2=-\left(
G_{c\theta}(0)
+f_2\right)/(2\alpha K_1),\label{k2}
\ee
which determines the crack path. In an isotropic material,
this condition reduces to the principle of local symmetry
since $G_{c\theta}$ vanishes trivially, and 
$f_2=0$.
The latter follows from the symmetry of the inner
phase-field solution for a propagating crack with $K_2=0$,
$\phi(x_1,x_2)=\phi(x_1,-x_2)$, which implies that
the product 
$\partial_1\phi\partial_2\phi$ in Eq.~(\ref{dissforces}) is
anti-symmetric. In an anisotropic material, however,
$\phi$ is generally not symmetrical and 
$f_2$ only vanishes in the zero velocity
limit where $G\rightarrow G_c$. 

The same procedure can
be repeated for pure antiplane shear where 
$K_3^*=K_3-b \mu A_2 \sqrt{L}\theta$ to linear
order in $\theta$ \cite{Sih1965} where $b$ is a numerical
constant, and hence $G_\theta(0)\sim K_3A_2 \sqrt{L}$.
One important difference with plane loading  
is the divergence of $G_\theta(0)$ 
with the crack extension length $L$. This divergence is also  
reflected in a $\sqrt{R}$ dependence of the integral
in Eq.~(\ref{f2}) on the radius $R$ of
the contour enclosing the tip. Since the only
natural cut off for this divergence is  
the system size, this result seems to imply
that the crack path cannot be  
predicted solely in terms of local conditions at the tip 
for mode III . We expect, however, this divergence to be cut off in
a real experiment by the process zone size $\xi$ due to the
irreversible nature of the fracture process. Namely,
fracture surfaces at a distance behind the tip larger
than $\xi$ should be essentially immobile, which implies
that $G_{\theta}(0)\sim K_3A_2 \sqrt{\xi}$ 
up to a numerical prefactor. 
Eq.~(\ref{x2}) then yields the local symmetry condition $A_2=0$
in the isotropic limit, 
where
the symmetry
of the phase-field profile
makes $f_2$ vanish, as explained above.
We will present elsewhere
numerical results that validate this condition
for mode III. We focus in the remainder of this
letter on plane strain.

We use a simple anisotropic extension of the 
phase-field model of Ref. \cite{KKL2001} with
an energy density
\be
{\cal E}=\kappa\left(|\nabla \phi|^2
+ \epsilon \partial_1\phi\partial_2\phi \right)/2
+g(\phi)\left({\cal E}_{strain}-{\cal E}_c\right),
\ee
where $u_{ij}=(\partial_i u_j+\partial_j u_i)/2$ is the strain
tensor and ${\cal E}_{\rm strain}\equiv \lambda u_{ii}^2/2
+\mu u_{ij}^2 $ is the strain energy. No asymmetry between  
dilation and compression is included since this is
not necessary here to test our predictions.
The broken state of the material becomes
energetically favored when ${\cal E}_{\rm strain}$ 
exceeds a threshold ${\cal E}_c$
and $g(\phi)=4\phi^3 - 3\phi^4$ is a monotonously increasing
function of $\phi$ that describes the softening of the elastic
energy at large strain. 
By repeating the analysis of Ref. \cite{KKL2001},
we obtain that 
$\gamma(\theta)=\gamma_0\sqrt{1-(\epsilon/2)\sin 2\theta}$
where $\gamma$ reduces to the isotropic surface energy 
of Ref. \cite{KKL2001} in the  
$\epsilon\rightarrow 0$ limit.  

We test our prediction for the initial angle $\theta$ of
a kink crack. Eq.~(\ref{eqmo}) is solved numerically using 
an Euler explicit scheme to integrate
the phase-field evolution and a successive
over relaxation (SOR) method to calculate the quasi-static 
displacement fields $u_1$ and $u_2$
at each time step. We used as initial 
condition a straight horizontal crack of length $2W$ 
centered in a strip of length $4W$ horizontally and $2W$ vertically,
with fixed values of $u_1$ and $u_2$ on the strip boundaries
that correspond to the  singular stress fields defined by Eq.~(\ref{km}) for 
prescribed values of $K_1$ and $K_2$.
We used $\lambda/\mu=1$ [$\alpha=3/(8\mu)$], a grid spacing 
$\Delta x_1=\Delta x_2=0.1\xi$, and $W=50 \xi$, 
where the process zone size 
$\xi\equiv \sqrt{\kappa/(\mu{\cal E}_c^2)}$. 
We checked that the results are 
independent of width and grid spacing.

\begin{figure}[t]
\includegraphics[width=5cm]{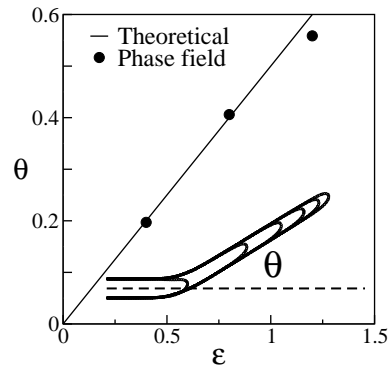}
\caption{Kink angle $\theta$ versus
surface energy anisotropy $\epsilon$
predicted as $\theta=\epsilon/2$
and extracted from phase-field simulations (filled circles)
for $G/G_c\approx 1.1$.
Inset:
phase-field simulation
for $\epsilon=1.2$ ($\phi=1/2$ contours are equally
spaced in time).
}
\label{compare}
\end{figure}

We have verified that the kink angle  
is well predicted by the local symmetry 
condition $K_2^*=0$
in the isotropic limit, which implies
that $\theta\approx -2K_2/K_1$. In the anisotropic
case, we choose $K_2=0$ and $G$ just slightly above
$G_c$ such that $f_2$ can be neglected in Eq.~(\ref{k2}). Substituting
$K_2^*\approx K_1\theta/2$ in Eq.~(\ref{k2}) and
using the fact that $(1-\nu^2)K_1^2/E\approx 2\gamma(0)$
for $G$ close to $G_c$, we obtain the prediction
for the kink angle $
\theta \approx -\gamma_\theta/\gamma\approx \epsilon/2$
which is strictly valid for $\theta\ll 1$ and $G\rightarrow G_c$.
This prediction is in good quantitative agreement
with the results of phase-field
simulations as shown in Fig.~\ref{compare}.

An interesting implication of our 
results for crystalline materials 
is that $|K_2|$ should exceed some threshold for a cleavage
crack to change direction. Using the expected cusp behavior
of the surface energy for small angle near a cleavage plane,
$\gamma(\theta)=\gamma_0(1+\delta|\theta|+\dots)$,
Eq.~(\ref{k2}) predicts that this threshold
is 
$E \gamma_0\delta/[(1-\nu^2)K_1]$ for $G\approx G_c$ 
since this equation cannot be satisfied
for any smaller value of $K_2$ for $\theta$ small. 
It should be hopefully possible to test this prediction 
experimentally as well as to explore the validity of 
this new condition on $K_2$ for curvilinear paths.
The extension of the present analysis to include
inertial effects and to three dimensions where
fracture paths are geometrically more complex
is an important future direction. Work along this line
is presently in progress.
 
We thank M. Adda-Bedia and J. B. Leblond for
valuable discussions. A.K. thanks the hospitality of 
ENS and support of DOE Grant 
No. DE-FG02-92ER45471.


\begin{thebibliography}{99}
\bibitem{refgen} K.~B.~Broberg, {\em Cracks and Fracture} (Academic Press, San 
Diego, 1999).

\bibitem{BarChe1961} G. I. Barenblatt and G. P. Cherepanov, 
PMM {\bf 25}, 1110 (1961) [J. Appl. Math. Mech. {\bf 25}, 1654 (1961)].
 

\bibitem{GolSal1974}  
R. V. Goldstein and R. L. Salganik, Int. J. Fract. {\bf 10}, 507 (1974)
and references therein.

\bibitem{CotRic1980} B. Cotterell and J. R. Rice, Int. J. Fract.
{\bf 16}, 155 (1980).

\bibitem{Mar2004} M. Marder,``Cracks Cleave Crystals'' (in press).

\bibitem{KKL2001} A. Karma, D. Kessler, and H. Levine, Phys. Rev. Lett. {\bf 87},
045501 (2001).

\bibitem{KarLob2004} A. Karma and A. Lobkovsky, Phys. Rev. Lett. {\bf 92},
245510 (2004).

\bibitem{HenLev2004} H. Henry and H. Levine, Phys. Rev. lett. {\bf 93},
105504 (2004).

\bibitem{Esh1975} J. D. Eshelby, J. Elast. {\bf 5}, 321 (1975),
and earlier references therein.

\bibitem{Guretal} M. E. Gurtin and P. Podio-Guidigli, J. Mech. Phys. Solids {\bf
44}, 1343 (1998).

\bibitem{Adaetal1999} M. Adda-Bedia {\it et al.},
Phys. Rev. E {\bf 60}, 2366 (1999). 

\bibitem{Ole2003} G. E. Oleaga, J. Mech. Phys. Solids {\bf 49}, 2273 (2001).

\bibitem{Ric1968} J. R. Rice, J. Appl. Mech. {\bf 35}, 379 (1968).

\bibitem{Her1951} C. Herring, \emph{The Physics of Powder Metallurgy},
ed. by W. E. Kingston (McGraw-Hill, New York, 1951), p. 143. 

\bibitem{AmeLeb1992} M. Amestoy and J. B. Leblond, Int. J. Solids Structures
{\bf 29}, 465 (1992), and earlier references therein.

\bibitem{Sih1965} G. C. Sih, J. Appl. Mech. {\bf 32}, 51 (1965).

\end{thebibliography}
\end{document}